\documentclass[twocolumn]{emulateapj} 

\newcommand{\beq}{\begin{equation}}
\newcommand{\eeq}{\end{equation}}
\newcommand{\bea}{\begin{eqnarray}}
\newcommand{\eea}{\end{eqnarray}}

\usepackage{amsmath}
\usepackage{amssymb}
\usepackage{textcomp}
\usepackage{color}

\shorttitle{H$_2$ and H$_2^+$ in the primordial gas}
\shortauthors{Coppola et al.}

\begin{document}

\title{Vibrational level population of H$_2$ and H$_2^+$ in the early Universe}

\author{Carla M. Coppola\altaffilmark{1,2}, Savino
Longo\altaffilmark{1,3}, Mario Capitelli\altaffilmark{1,3}, Francesco
Palla\altaffilmark{4}, Daniele Galli\altaffilmark{4}}

\altaffiltext{1}{Universit\`a degli Studi di Bari, Dipartimento di
Chimica, Via Orabona 4, I-70126 Bari, Italy}
\altaffiltext{2}{Department of Physics and Astronomy, University
College London, Gower Street, London WC1E 6BT}
\altaffiltext{3}{IMIP-CNR, Section of Bari, via Amendola 122/D, I-70126
Bari, Italy} \altaffiltext{4}{INAF-Osservatorio Astrofisico di Arcetri,
Largo E.~Fermi 5, I-50125 Firenze, Italy}

\begin{abstract}

We formulate a vibrationally resolved kinetics for molecular hydrogen
and its cation in the primordial Universe chemistry.  Formation,
destruction and relaxation processes for each
vibrational level are studied and included as
chemical pathways of the present model.
The fractional abundance of
each vibrational level as a function of the redshift is given: a strong
deviation from the Boltzmann distribution is found at low $z$. A
discussion of the results is provided, also evaluating the effects of
relaxation processes on the level populations. 
Analytical fits for some LTE rate
coefficients are given in the Appendix.


\end{abstract}


\section{Introduction}

The primordial gas after recombination was composed of neutral
hydrogen and helium atoms with traces of electrons, deuterium and
lithium, exposed to the radiation field of the cosmic
background radiation (hereafter CBR), and in adiabatic expansion. In these
conditions, only a limited number of gas-phase reactions
was possible. The first molecules started to form only
ten thousand years after the Big Bang, at a resdhift $z\approx 2000$,
reaching a freeze-out abundance at $z\approx 100$: about $10^{-6}$ for
H$_2$, followed by HD, H$_2^+$, HeH$^+$ and several less abundant species
(see Lepp et al.~2002 for a review).

Despite the simple chemical composition, the kinetics of elementary
processes in the early Universe is very complex, due to the presence of
a large number of quantum states for atoms and molecules involved in
the network processes, including the interaction with
the CBR. Detailed calculations of molecule formation
in the early Universe have been recently performed by
Hirata \& Padmanabhan~(2006, hereafter HP06), Puy \& Signore~(2007),
Glover \& Jappsen~(2007), Schleicher et al.~(2008,
hereafter S08), and Vonlanthen et al.~(2009,
hereafter V09).  In addition to these
studies, comprehensive collections of analytic fits of chemical
reaction rates were also given by Anninos \& Norman~(1996), Stancil et
al.~(1998), Galli \& Palla~(1998, hereafter GP98) and,
 more recently, by Glover \& Abel~(2008).

In studies of primordial chemistry, the rovibrational
levels manifold of the molecules has been usually
ignored, mainly because of the lack of state-to-state resolved cross
sections (or, equivalently, reaction rates) of the relevant chemical
processes.  Early estimates of the vibrational distribution function
of H$_2$ attempted by means of physical
order-of-magnitude arguments (Khersonskii~1982) led to the conclusion
that the vibrationally excited states of H$_2$ are rapidly deactivated
through radiative decay. This is because the H-H$_2$ collision time is
much longer than the radiative lifetime of vibrational
levels, so that the population of the vibrational levels is essentially
determined by the Einstein coefficients and the probability of
formation of vibrationally excited $\mathrm{H_2}$ molecules.  HP06 were
the first to compute in detail the vibrational level populations of
H$_2^+$.  However, they did not consider collisional processes inducing
vibrational relaxation.

Because of these simplifications, chemical reaction rates are usually
evaluated under the hypothesis of local thermal equilibrium (hereafter
LTE). However, non-equilibrium distribution function can significantly
affect the results of such calculations, as shown explicitly by
Capitelli et al.~(2007) for the dissociative attachment process of
H$_2$.  This evidence suggests that the evaluation of the exact
vibrational distribution function is necessary to model the plasma
kinetics, as pointed out by Lepp et.~(2002). The relatively simple
internal structure of atomic and molecular hydrogen, and its molecular
cation H$_2^+$, makes possible such a {\it
state-to-state} approach at least for these species.

In the present work, we follow for the first time the
chemical composition of the primordial gas, determining the population
of vibrational levels of H$_2$ and H$_2^+$, adopting a detailed network
of formation and destruction channels and relaxation
processes, such as radiative and atom/ion-molecule
collisions. We ignore any contribution due to subsequent astrophysical
processes that affect the chemical composition of the gas such as
primordial massive supernovae, which, besides providing heavy elements,
 may also contribute to the formation of H$_2$
in the early Universe (Cherchneff \& Lilly~2008).

The paper is organized as follows: in Section~2 we discuss the relevant
vibrationally resolved reactions and we compute state-to-state reaction
rate coefficients; in Section~3 we describe our chemical network; in
Section~4 we present our results for the evolution of the vibrational
level populations of H$_2$ and H$_2^+$; finally, in Section~5 we
summarize our conclusions. In the Appendix, we give a
list of fitting formulae for specific rate coefficients
under LTE conditions.


\section{Vibrationally resolved reaction rates for $\mathrm{H_2}$ and
$\mathrm{H_2^+}$}
In a hydrogen plasma,
H$_2$ and H$_2^+$ form
by various collision processes, generally in vibrationally
excited states. For example, in the case of H$_2^+(v)$, the electron
impact ionization process produces
a known population of vibrational states 
given by the corresponding Franck-Condon factors
for the transitions. For other formation processes,
H$_2^+(v)$ is created with different populations of vibrational
levels. Additionally, inelastic collisions with other plasma constituents
strongly affect the evolution and thermalization of vibrational
levels of H$_2$ and H$_2^+$.
This fact underlines the importance
of such a vibrationally resolved kinetics, since a priori assumptions and approximations could
fail in evaluating the vibrational level distribution.

Typically, collisional processes in a low-temperature plasma involve
all molecular degrees of freedom (electronic, vibrational and
rotational); however, in order to model the kinetics of vibrational
levels of H$_2$ and H$_2^+$, rotational equilibrium
has been assumed. 

For processes not involving H$_2$ and H$_2^+$, we have adopted the
chemical network recently developed by S08, updating the value of the
rate coefficient for the H$^-$ photodetachment to include the
contribution of non-thermal photons due to cosmological H and He
recombination (see HP06). The corresponding rate
coefficient for this process is given in Table~\ref{fits}. For H$_2$ and H$_2^+$, we computed
state-to-state rate coefficients from available cross sections,
including all vibrational levels of their fundamental electronic states (15 and 19 states, respectively).


A list of the chemical processes included in our calculation is
shown in Table~\ref{vibsolvedchemprocess}.  Each process will be
described separately in the following subsections.

\begin{table}
\caption{Vibrationally resolved reactions included in this work}
\begin{tabular*}{\columnwidth}{ll}
\hline
Chemical process & Reference \\
\hline
& \\
$1] ~~\mathrm{H}+\mathrm{H_2}(v) \rightarrow \mathrm{H}+\mathrm{H_2}(v^\prime)$      & (1) \\
$2] ~~\mathrm{H}+\mathrm{H_2}(v) \rightarrow \mathrm{H}+\mathrm{H}+\mathrm{H}$       & (1) \\
$3] ~~\mathrm{H}+\mathrm{H^-} \rightarrow \mathrm{H_2}(v) + \mathrm{e^-}$            & (2) \\
$4] ~~\mathrm{H}+\mathrm{H^+} \rightarrow \mathrm{H_2^+}(v)+h\nu$     & detailed balance from (6) \\
$5] ~~\mathrm{H_2^+}(v)+\mathrm{H} \rightarrow \mathrm{H_2}(v^\prime)+\mathrm{H^+}$  & (3) \\
$6] ~~\mathrm{H_2}(v)+\mathrm{H^+} \rightarrow \mathrm{H_2^+}(v^\prime)+\mathrm{H}$  & (3) \\
$7] ~~\mathrm{H_2^+}(v)+\mathrm{H} \rightarrow \mathrm{H_2^+}(v^\prime)+\mathrm{H}$  & (3) \\
$8] ~~\mathrm{H_2}(v)+\mathrm{H^+} \rightarrow \mathrm{H_2}(v^\prime)+\mathrm{H^+}$  & (3) \\
$9] ~~\mathrm{H_2^+}(v)+\mathrm{H} \rightarrow \mathrm{H^+}+\mathrm{H}+\mathrm{H}$   & (4) \\
$10] ~~\mathrm{H_2}(v)+\mathrm{H^+} \rightarrow \mathrm{H}+\mathrm{H}+\mathrm{H^+}$  & (4) \\
$11] ~~\mathrm{H_2}(v)+h\nu \rightarrow \mathrm{H_2^+}(v^\prime)+\mathrm{e^-}$        & (5) \\
$12] ~~\mathrm{H_2^+}(v) + h\nu \rightarrow \mathrm{H}+\mathrm{H^+}$                 & (6) \\
$13] ~~\mathrm{H_2}(v)+h\nu \rightarrow \mathrm{H}+\mathrm{H}$         &  direct + indirect (7)  \\
$14] ~~\mathrm{H_2^+}(v)+\mathrm{e^-} \rightarrow \mathrm{H}+\mathrm{H}$             & (8) \\
$15] ~~\mathrm{H_2}(v)+\mathrm{e^-} \rightarrow \mathrm{H^-}+\mathrm{H}$             & (9) \\
$16] ~~\mathrm{H_2}(v) + \mathrm{e^-} \rightarrow \mathrm{H_2}(v^\prime)+\mathrm{e^-}+h\nu$  & (9) \\
$17] ~~\mathrm{H_2}(v) \rightarrow \mathrm{H_2}(v^\prime)+h\nu$                      & (10) \\
$18] ~~\mathrm{H_2^+}(v) \rightarrow \mathrm{H_2^+}(v^\prime)+h\nu$                  & (11) \\
$19] ~~\mathrm{HeH^+} + \mathrm{H} \rightarrow \mathrm{H_2^+}(v)+\mathrm{He}$        & (12) \\
& \\
\hline
\end{tabular*}
{References: 
(1) Esposito et al.~(1999, 2001); (2) \u{C}\'{\i}\v{z}ek et al.~(1998);
(3) Krsti\'{c} et al.~(1999, 2002), Krsti\'{c}~(2002); (4) Krsti\'{c}~(2003); (5) Flannery et
al.~(1977); (6) Dunn~(1968); (7) Allison \& Dalgarno~(1969, 1970), Dalgarno
\& Stephens~(1970); (8) Takagi~(2002); (9) Celiberto et al.~(2001); (10)
Wolniewicz et al.~(1998); (11) Posen et al.~(1983); (12) Linder et al.~(1995) (see S08).}
\label{vibsolvedchemprocess}          
\end{table}

\subsection{Associative detachment}
\label{subcizek}

The importance of the reaction of associative detachment for H$^-$,
\[
\mathrm{H}+\mathrm{H^-} \rightarrow \mathrm{H_2}(v)+\mathrm{e^-},
\]
was first pointed out by Dalgarno (quoted by Pagel~1959). 
Various studies have been published on the vibrationally resolved cross
sections and/or rate coefficients for this process (Bieniek \&
Dalgarno~1979, Launay et al.~1991), underlying the importance of
this channel of H$_2$ formation (e.g. Dalgarno~2005;
Flower~2000).

\begin{figure}
\includegraphics[width=6cm,angle=-90]{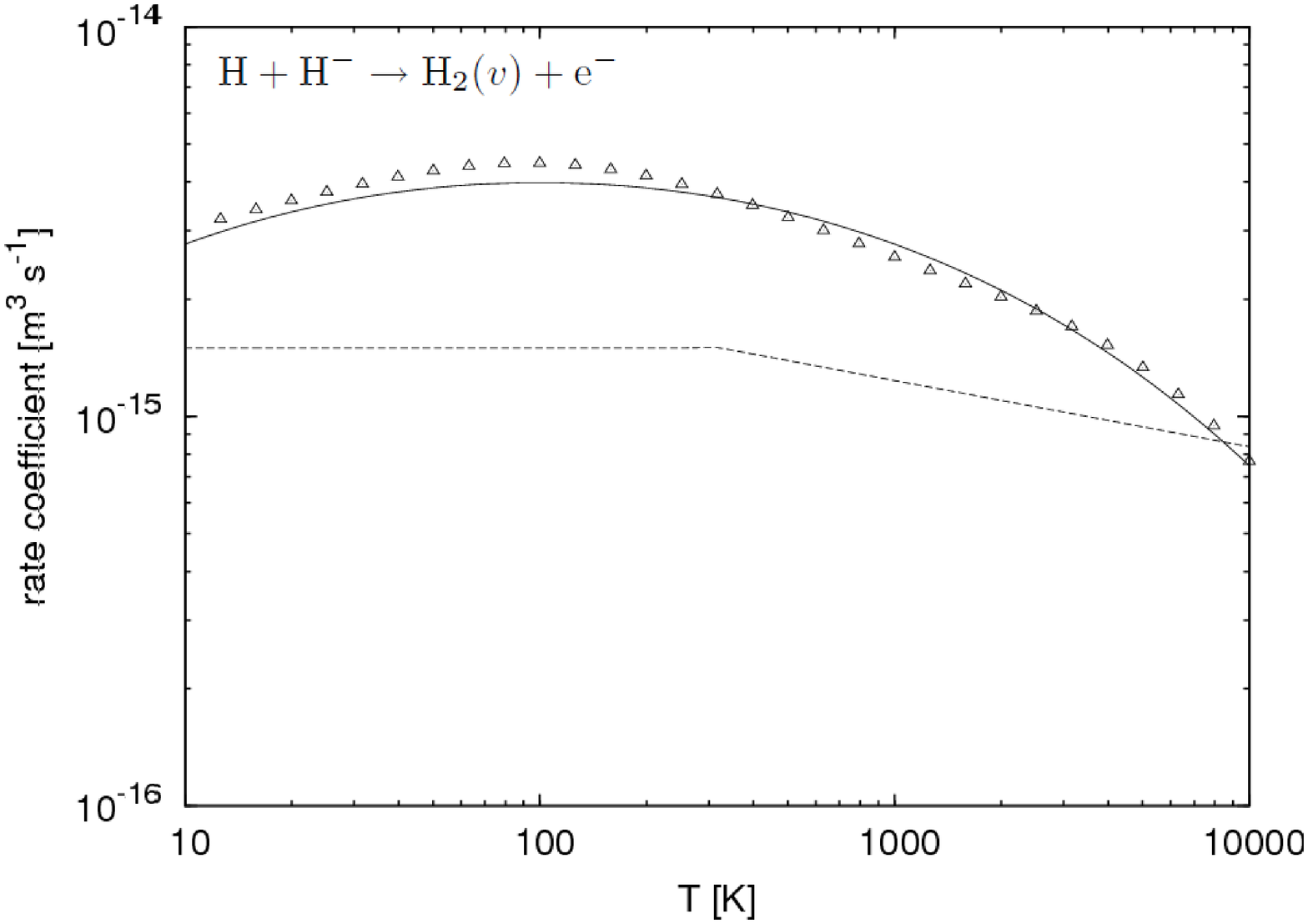} 
\caption{H--H$^-$ associative detachment rate coefficient.
{\it Triangles}, total rate coefficient obtained from the cross sections of 
\u{C}\'{\i}\v{z}ek et al.~(1998), according to Eq.~(\ref{ad_tot}); {\it
dashed curve}, fit by GP98 of the total rate coefficient from Launay et al.~(1991); {\it
solid curve}, fit reported in Table~\ref{fits}.}
\label{cizekcomp} 
\end{figure}

Our calculations are based on the
detailed data by
\u{C}\'{\i}\v{z}ek et al.~(1998).  These authors developed an improved
nonlocal resonance model for the description of the nuclear dynamics of
the H$_2^-$ collision complex, and calculated both associative
detachment and dissociative attachment cross sections. They provide
data for different values of the angular momentum $J$ of the H$^-$+H
collision\footnote{Data in electronic format are available at {\tt
http://utf.mff.cuni.cz/$^\sim$cizek/AD$\_$H2/}.}. For each initial
angular momentum, the H$_2$ molecule can be produced
in states with the
rotational quantum number equal to $J+1$ or $J-1$ due to selection
rules; states with the same $J$ are not allowed by symmetry conditions
on the total wavefunction and higher changes in $J$ are neglected in
their model. The total rate coefficient, obtained as a sum of the
vibrationally resolved rate coefficients,
\beq 
k_{\rm tot}(T_{\rm gas})=\sum_v k_v(T_{\rm gas}),
\label{ad_tot}
\eeq 
is shown in Fig.~\ref{cizekcomp}, compared with
the rate coefficient computed by Launay et al.~(1991),
as fitted by GP98. The new fit is given in Table~\ref{fits}.
Using this rate coefficient in a 
non-vibrationally resolved chemical network of the primordial Universe
has a negligible impact on the H$_2$ abundance. Indeed, this reaction is also one of the most important H$^-$ loss channel.
Therefore increasing the associative detachment rate coefficient cannot lead to a production rate of H$_2$ faster than the H$^-$
 production rate. For the same reasons,
H$^-$ fraction is reduced accordingly, by a factor of 4.

\subsection{Photodissociation of $\mathrm{H_2^+}(v)$}

\begin{figure}
\includegraphics[width=6cm,angle=-90]{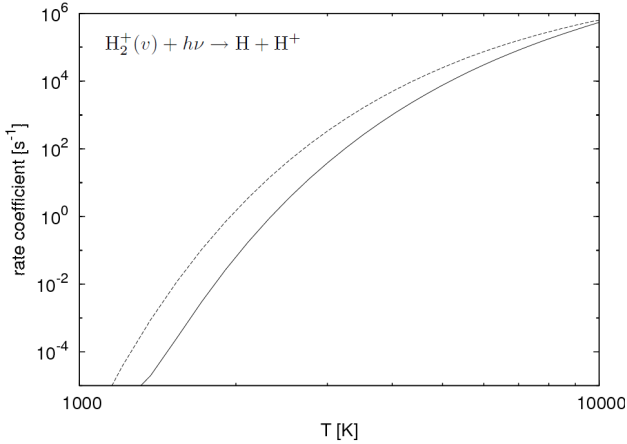} 
\caption{H$_2^+$ photodissociation rate coefficient. {\it Solid curve},
total rate in LTE, summed over all possible initial,
from the cross sections by Dunn~1968; {\it dashed curve},
fit by GP98 based on data by Argyros~(1974) and Stancil~(1994).}
\label{h2pphotodissociation} 
\end{figure}

The photodissociation of H$_2^+(v)$, 
\[
\mathrm{H_2^+}(v)+h\nu\rightarrow \mathrm{H}+\mathrm{H^+},
\]
is a typical example of direct process of photodestruction. Lebedev et
al.~(2000, 2003) provided a significant analysis of bound-free
(including also photodissociation) and free-free transitions for
H$_2^+(v)$, limited however to sums on the rotational and vibrational
quantum numbers. In this work, we have adopted the vibrationally
resolved cross sections calculated by Dunn~(1968),
evaluated fixing the rotational quantum number $J=1$. 
The comparison between the LTE rate coefficient obtained using the vibrationally resolved data by Dunn
and the fit by GP98 of data by
Argyros~(1974) for
$2500$~K$<T_{\rm rad}<26000$~K and by Stancil~(1994) for
$3150~$K$<T_{\rm rad}<25200$~K is shown in Fig.~\ref{h2pphotodissociation}.

\subsection{Radiative association of $\mathrm{H}$ and $\mathrm{H^+}$}
\label{radasssection}
\begin{figure}
\includegraphics[width=6cm,angle=-90]{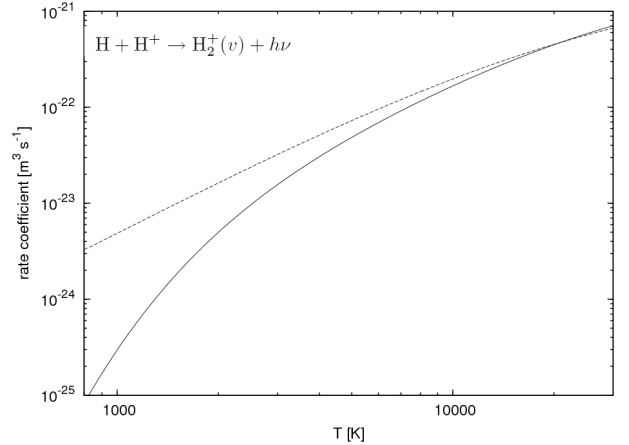} 
\caption{H--H$^+$ radiative association rate coefficient.  {\it Solid
curve}, total rate coefficient obtained using the detailed balance
principle from the photodissociation data by Dunn~(1968), according to
Eq.~(\ref{rcradass}); {\it dashed curve}, fit by GP98 of the total rate
coefficient from Ramaker \& Peek~(1976).}
\label{radassfigure} 
\end{figure}

The reaction of
radiative association of H and H$^+$,
\[
\mathrm{H}+\mathrm{H^+} \rightarrow \mathrm{H_2^+}(v)+h\nu,
\]
has been the subject of several studies
(e.g., Ramaker \& Peek~1976, Shapiro \& Kang~1987,
Stancil et al.~1993). However, no vibrationally resolved data are
available in literature. For this reason, we have applied
the principle of detailed balance to the inverse H$_2^+(v)$ photodissociation
reaction,
\beq
k_v^{\rm rad. ass.}=
\frac{Z_{{\rm H}_2^+(v)}}{Z_{\rm H} Z_{{\rm H}^+}}
e^{h\nu_v/kT_{\rm rad}}\,k_v^{\rm ph.}
\label{rcradass}
\eeq
where $T_{\rm rad}$ is the temperature of the CBR,
$h\nu_v$ is the thermal threshold for the $v$-th level.
The partition functions $Z$ of reagents and product
are

\[
Z=\frac{(2\pi kT_{\rm gas})^{3/2}}{h^3} 
\left\{ \begin{array}{l}
m_{\rm H}^{3/2} g_{\rm H} Q_{\rm H}~~~\mbox{for H},\\
m_{{\rm H}^+}^{3/2} g_{{\rm H}^+}~~~\mbox{for H$^+$},\\
m_{{\rm H}_2^+}^{3/2} g_{{\rm H}_2^+}Q_{{\rm H}_2^+}Z_{\rm rot}(v,T_{\rm gas})~\mbox{for H$_2^+(v)$},\\
\end{array} \right.
\] 
where $g_{\rm H}$, $g_{{\rm H}^+}$ and $g_{{\rm H}_2^+}$ are the
multiplicities of the nuclear spin contributions for H, H$^+$ and
H$_2^+$, equal to $g_{\rm H}=g_{{\rm H}^+}=2\times (1/2)+1=2$ and
$g_{{\rm H}_2^+}=4$; $Q_{\rm H}$ and $Q_{{\rm H}_2^+}$ are the electron
spin degeneracies, equal to 2 both for H and H$_2^+$ under the
hypothesis that only their ground electronic states contribute to the
partition function and $Z_{\rm rot}(v,T_{\rm gas})$ is the rotational
partition function, that takes into account the rotational structure of
each vibrational level of the cation hydrogen molecule,
given by
\[
Z_{\rm rot}(v,T_{\rm gas})=\sum_J g_J e^{-(E_0-E_{vJ})/kT_{\rm gas}},
\]
where $E_{vJ}$ is the binding energy of the $(v,J)$-th level and $E_0$
the binding energy of the lowest rovibrational level.  
The symbol $g_J$ represents the multiplicity of each electronic-rotational level,
\[
g_J=\left\{ 
\begin{array}{l}
\frac{1}{4}(2J+1) \qquad \mbox{for $J$ even}, \\
\frac{3}{4}(2J+1) \qquad \mbox{for $J$ odd}. 
\end{array} 
\right.
\]
A comparison between the present LTE calculation and the
fit by GP98 of data by Ramaker \& Peek~(1976) is shown
in Fig.~\ref{radassfigure}.

\subsection{Charge transfer, collisional excitation, and dissociation}

The H$^+$--H$_2$ and the H--H$_2^+$ systems represent one of the most
important collision complexes in the hydrogen plasma, called in short
the ${\rm H}_3^+$ collision system. In addition to elastic 
collisions H$_2$--H$^+$ and H$_2^+$--H, the relevant reactions are:
({\it i}\/) charge transfer,
\[
\begin{array}{l}
\mathrm{H_2}(v)+\mathrm{H}^+\rightarrow \mathrm{H_2^+}(v^\prime)+\mathrm{H}~~({\it a}\/),\\
\mathrm{H_2^+}(v)+\mathrm{H}\rightarrow \mathrm{H_2}(v^\prime)+\mathrm{H}^+~~({\it b}\/),\\
\end{array} 
\]
and ({\it ii}\/) collisional dissociation,
\[
\begin{array}{l}
\mathrm{H_2}(v)+\mathrm{H}^+\rightarrow \mathrm{H}+\mathrm{H}+\mathrm{H}^+~~({\it a}\/),\\
\mathrm{H_2^+}(v)+\mathrm{H}\rightarrow \mathrm{H}+\mathrm{H}+\mathrm{H}^+~~({\it b}\/).\\
\end{array} 
\]

\begin{figure}
\includegraphics[width=9cm]{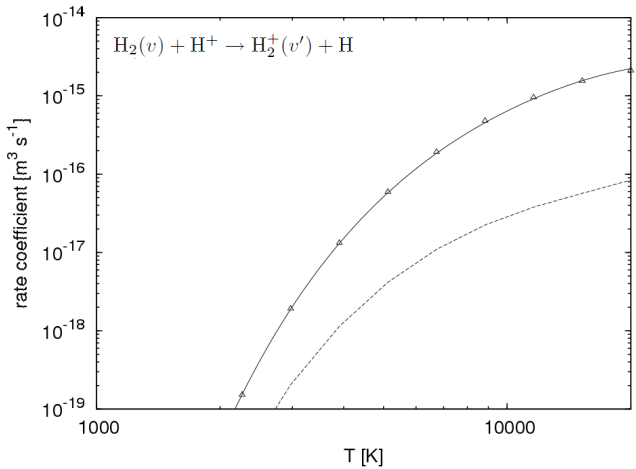} 
\caption{H$_2$--H$^+$ charge transfer rate coefficient. {\it Triangles}, total rate in LTE, summed over all possible initial and final
states, from the cross sections of Krsti\'c~(2002) and Krsti\'c et
al.~(2002, 2003); {\it dashed curve}, fit by Savin et al.~(2004) for
H$_2(v=0)$; {\it
solid curve}, fit listed in Table~\ref{fits}.}
\label{cth2hp} 
\end{figure}

\begin{figure}
\includegraphics[width=9cm]{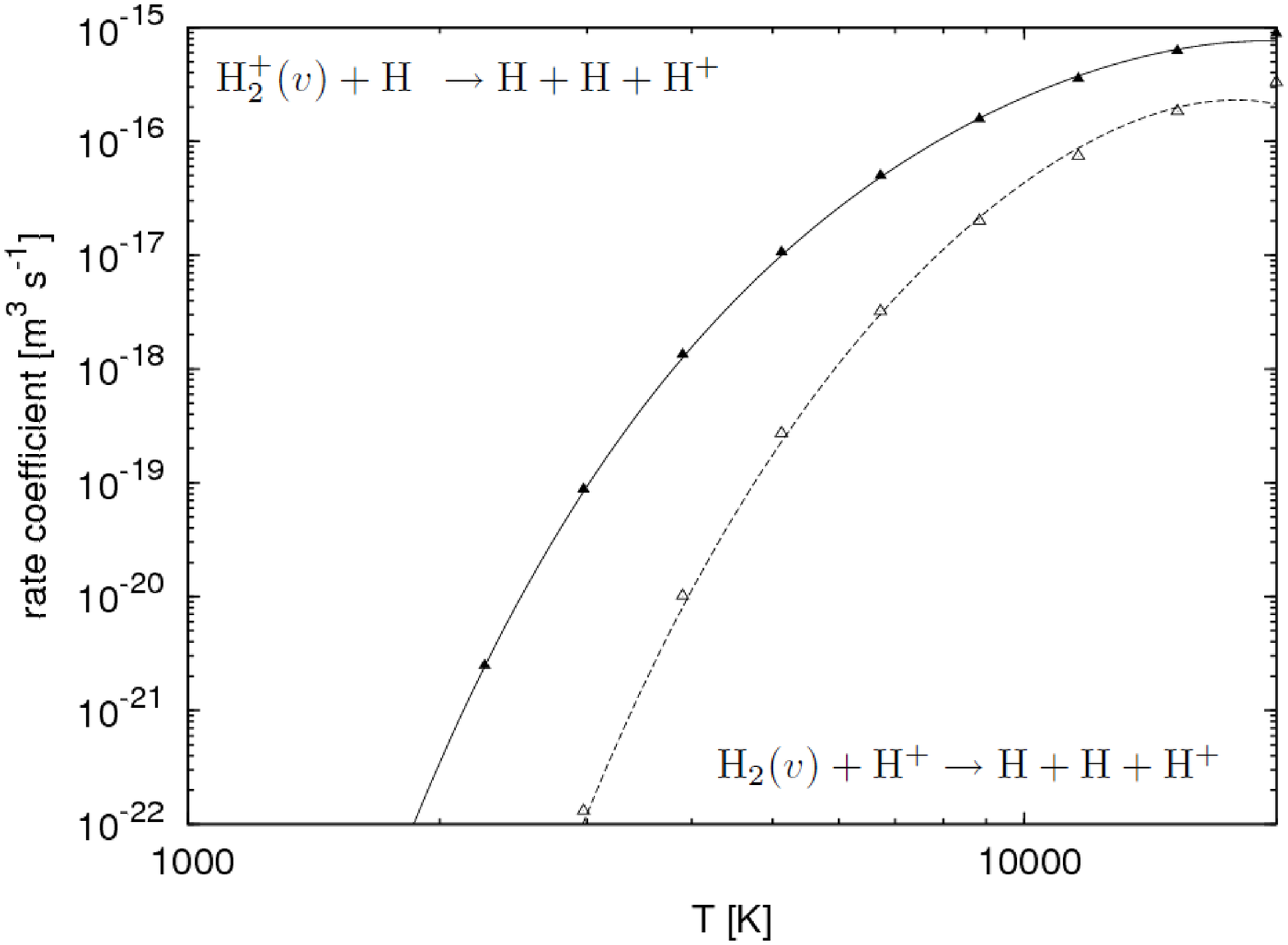} 
\caption{H$_2^+$--H ({\it filled triangles}) and H$_2$--H$^+$ ({\it
open triangles}) total dissociation rate coefficients in LTE, summed
over all possible initial and final states, from the cross sections of
Krsti\'c~(2002) and Krsti\'c et al.~(2002, 2003). The fits
listed in Table \ref{fits} are also shown as solid and
dashed lines, respectively.}
\label{dissh3+} 
\end{figure}

The most complete theoretical work describing these chemical reaction
pathways has been recently developed by Krsti\'{c} and collaborators
(Krsti\'c~2002, 2003, 2005; Krsti\'c et al.~1999, 2002) within the
\emph{infinite order sudden approximation} method (IOSA)\footnote{Data
in electronic form are available at {\tt
http://www-cfadc.phy.ornl.gov/astro/ps/data/home.html}.}.  As described
by Krsti\'{c} et al.~(2002), charge transfer ({\it a}\/) is endoergic
for $v\le 3$, although with low threshold energies, and is strictly
competitive with the vibrational excitation channel, especially for
those states than can overcome the kinetic barrier ($v\ge 4$). For
higher vibrational levels, the process is exoergic, as the process of
charge transfer ({\it b}), for all $v$. The LTE rate
coefficient for the charge exchange reaction ({\it a}\/) is shown in
Fig.~\ref{cth2hp}, and compared with the fit by Savin et al. (2004) for H$_2(v=0)$. For the 
reaction ({\it b}\/), cross-sections have been extrapolated at low
energies using a Langevin-type power law; the resulting rate
coefficients have been scaled in order to obtain an LTE value
corresponding to the experimental measurement by Karpas et al.~1979 at
$T_{gas}=300~$K. The rate coefficients for the collisional
dissociation reactions ({\it a}\/) and ({\t b}\/) are shown in
Fig.~\ref{dissh3+}. The fits are given in Table~\ref{fits}.

\subsection{$\mathrm{H_2}/\mathrm{H}$
vibrational-translational transfer}

Elastic collisions among hydrogen atoms and molecules deeply modify the
initial vibrational distribution function of molecular components in a
hydrogen plasma. They are generally classified as V-T
(vibrational-translational) and V-V (vibrational-vibrational)
processes.  In the former case, energy transfer between the
translational degree of freedom of the colliding atom and the
vibrational state of the target molecule occurs; the latter represents
the energy exchange between the vibrational manifolds of the colliding
molecules. In both cases, multi-quantum transitions can occur, coupling
the entire set of vibrational levels.  Recently,
Krsti\'{c} et al.~(2002) have calculated V-T cross sections for
collisions between H and H$_2$,
\[
\mathrm{H}+\mathrm{H_2}(v) \rightarrow \mathrm{H}+\mathrm{H_2}(v^\prime),
\]
using the quantum mechanical IOSA method. However, while
data are available for the collision system
H$^+$--H$_2(v)$, no state-to-state information on collisions among
hydrogen atoms and molecules is available for all
$(v,v')$ pairs. The V-T rate coefficients used in the present work have
been obtained from cross sections computed by Esposito et
al.~(1999, 2001) using the \emph{quasiclassical
trajectory} (QCT) method.

Collisions of H$_2$ with He atoms have been studied by
e.g. Clark~(1997), Flower et al.~(1998), Balkrishnan et al.~(1999), Lee
et al.~(2005), but no data for each $(v,v^\prime)$
pair are available. In this work we have neglected V-T
processes with He. This assumption is based on the observation that
vibrating nuclei in H$_2$ are lighter than the colliding He atom;
moreover, the relative abundance of He is small:  for these reasons,
the V-T transfer with He can be considered less
efficient than V-T transfer with H.
Also molecule-molecule collisions have been neglected,
owing to the small fractional abundance of H$_2$ with respect to H.

\subsection{Collisional dissociation of $\mathrm{H_2}$}

\begin{figure}
\includegraphics[width=9cm]{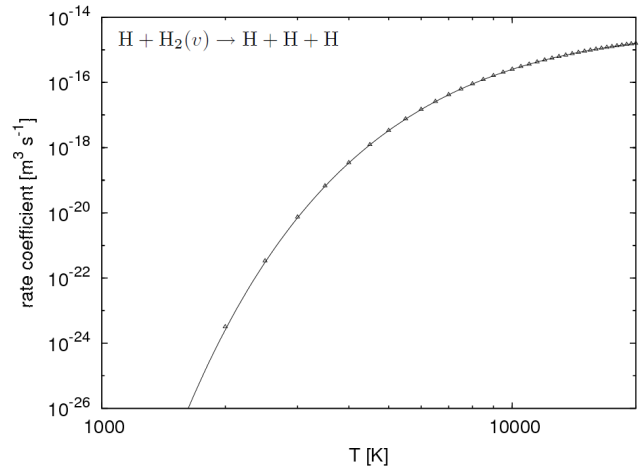} 
\caption{H$_2$ dissociation by H impact: 
{\it Triangles}: rate coefficient computed from the cross sections of 
Esposito et al.~(1999); {\it solid line}: fit reported in Table \ref{fits}.}
\label{dissfab} 
\end{figure}

The cross sections for the dissociation of H$_2$ induced by collisions
with H,
\[
\mathrm{H}+\mathrm{H_2}(v)\rightarrow \mathrm{H}+\mathrm{H}+\mathrm{H},
\]
have been computed by Esposito et al.~(1999, 2001) with the QCT method.
The whole vibrational manifold of H$_2$ has been explored. The total
LTE rate coefficient is shown in Fig.~\ref{dissfab},
and the fit is given in Table~\ref{fits}.

\subsection{Photoionization of $\mathrm{H_2}$}

\begin{figure}
\includegraphics[width=9cm]{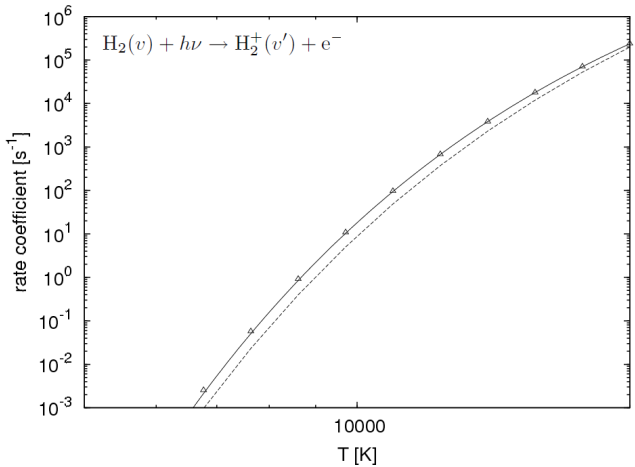} 
\caption{H$_2$ photoionization rate coefficient. {\it Triangles},
total rate in LTE, summed over all possible initial and final states,
from the cross sections by Flannery et al.~(1977); {\it dashed curve},
fit by GP98 based on data by ONeil \& Reinhardt~(1978); {\it solid curve}, fit given in Table \ref{fits}.}
\label{photoionization} 
\end{figure}

Ionization of H$_2$ can proceed through collisions
with particles and photons. Data on ionization caused by electron
impact are described by Liu \&
Shemansky~(2004), and compared with photoionization data. Because
threshold energies for the electron-impact ionization cross sections
are much higher than thermal energies available at the redshift range
considered in the present work ($E>16$~eV vs. $E<2$~eV), we have
considered only photoionization of H$_2$,
\[
\mathrm{H_2}(v)+h\nu\rightarrow \mathrm{H_2^+}(v^\prime)+\mathrm{e^-}.
\]
Several theoretical calculations of H$_2$ photoionization have been
proposed during the years (Lewis Ford et al.~1975, ONeil \&
Reinhardt~1978). This chemical reaction couples each vibrational level
of H$_2$ with each vibrational level of H$_2^+$. Available data on
vibrationally resolved $(v,v^\prime)$ cross sections have been
published by Flannery et al.~(1977), following the theoretical and
computational procedures described by Tai \& Flannery~(1977).
Fig.~\ref{photoionization} shows the comparison between the total
photoionization rate coefficient calculated in the hypothesis of LTE
using the cross sections by Flannery et al.~(1977), whose fit is given in Table \ref{fits},
 and the fit by GP98
based on data by ONeil \& Reinhardt~(1978).

\subsection{Photodissociation of $\mathrm{H_2}$} 

Photodissociation of H$_2$,
\[
\mathrm{H_2}(v)+h\nu\rightarrow \mathrm{H}+\mathrm{H},
\]
can occur both directly (absorption from a lower level into the continuum)
or indirectly (the so-called ``Solomon process''), a process consisting in the
absortpion into an individual level of a bound upper state, followed by 
dissociation. In the present work, we considered both
destruction pathways. For the direct process,
vibrationally resolved rate coefficients are given by
\beq
k_v^{\rm ph.}(T_{\rm rad})=4\pi\int_0^\infty \frac{\sigma_v^{\rm ph.}(\nu)}{h\nu} 
J_{\nu}(T_{\rm rad})\,d\nu,
\label{rcrad}
\eeq
where the cross sections $\sigma_v^{\rm ph.}$ are taken from the work
by Allison \& Dalgarno~(1969).  More recently, new calculations of
photodissociation cross sections have been published by
Glass-Maujean~(1986) and Zucker \& Eyler~(1986), including excited
electronic states; however, no significant changes have been introduced
for the Lyman and Werner systems.  A thorough analysis of the direct
process with related fits and reaction rates for the kinetics of the
early Universe is given in Coppola et al.~(2011).

\begin{figure}
\includegraphics[width=9cm]{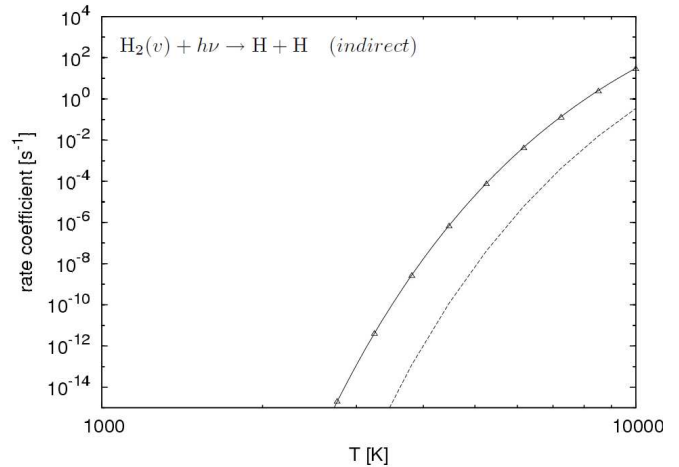} 
\caption{H$_2$ indirect photodissociation rate coefficient. {\it Triangles},
total rate in LTE, both considering the Lyman and the Werner system, obtained using Eq.~(\ref{rcrad2}) and
data by Dalgarno \&
Stephens~(1970) and Allison \& Dalgarno~(1970); {\it dashed curve},
fit by S08 on Glover \& Jappsen~(2007) data; {\it solid curve}, fit given in Table \ref{fits}.}
\label{solomon} 
\end{figure}

For the calculation of the rate coefficient of
the indirect process, we have used the discrete version of Eq.~(\ref{rcrad}) 
\beq
k_v^{\rm ph.}(T_{\rm rad})=\frac{\pi e^2}{m_{\rm e}c^2} \lambda_{uv}^2 f_{uv} 
\epsilon_u J_{\lambda_{uv}}(T_{\rm rad}),  
\label{rcrad2}
\eeq
where $u$ is the vibrational level of both Lyman and Werner
electronically excited intermediate systems, whose dissociation
efficiencies $\epsilon_u$ have been taken from Dalgarno \&
Stephens~(1970). The oscillator strengths $f_{uv}$ for the same
molecular states have been calculated by Allison \& Dalgarno~(1970),
and the vibrational energies of both Lyman and Werner states by Fantz
\& W\"{u}nderlich~(2006). In Fig.~\ref{solomon} we show the
comparison between the fit of the rate coefficient by S08 of the 
data by Glover \& Jappsen~(2007)
and the total LTE rate coefficient obtained using Eq.~(\ref{rcrad2}).
The present calculation includes both Lyman and Werner
systems, and the contribution of the total vibrational manifold has
been considered.

\subsection{Dissociative excitation and dissociative recombination of
$\mathrm{H_2^+}$}

The fate of the dissociative scattering processes between electrons and
H$_2^+$ strictly depends on the energy of the collision.
The effectiveness of these chemical reactions is connected to the energy
thresholds of the vibrational states of the intermediate excited states:
at high collision energies, the favored dissociative channel 
is dissociative excitation (DE):
\[
\mathrm{H_2^+}(v)+\mathrm{e}^-\rightarrow
\left\{ \begin{array}{l}
\mathrm{H_2^+}(2p\sigma_u)+\mathrm{e}^- \\
\mathrm{H_2^+}(2p\pi_u)+\mathrm{e}^- \\
\end{array} 
\right\}
\rightarrow
 \mathrm{H}+\mathrm{H}^++\mathrm{e}^-.\\
\]
However, experimental evidence (Yousif \& Mitchell~1995) has shown that
large DE cross sections are operative at energies $\sim 0.01$~eV. For
this reason, other reaction pathways become important at
these energies, involving electron capture into
excited electronic states of H$_2$. In particular,
$\mathrm{H_2}(^1\Sigma^+_g)$ or auto-ionizing dissociative Rydberg
states lying below the $\mathrm{H_2^+}(2p\sigma_u)$, both auto-ionizing
in the continuum of $\mathrm{H_2^+}$, have been considered in the
dynamical mechanisms of dissociative excitation.

At low energies, DE processes are competitive with
dissociative recombination (DR), in which the incident electron is
temporally captured into the Rydberg state by transferring its kinetic
energy to vibrational motion. The unstable molecule can dissociate or go
back to the ionizing state by an elastic or vibrational
transition.  This mechanism is sometimes called ``indirect process'' of
DR. In the direct process, excited hydrogen molecules
dissociate in the vibrational continuum of H$_2$. In this case,
two H atoms are produced, one in the ground electronic
state, the other in states with principal quantum number $n \geq 2$:
\[
\mathrm{H_2^+}(v)+\mathrm{e}^-\rightarrow
\mathrm{H}+\mathrm{H}(n\ge2).  
\]

\begin{figure}
\includegraphics[width=6cm,angle=90]{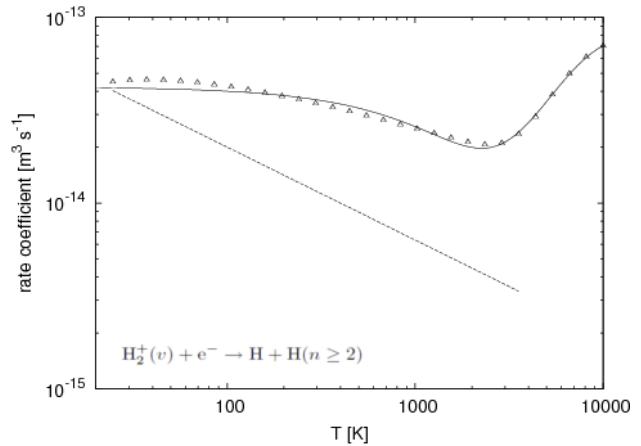} 
\caption{H$_2^+$ dissociative recombination rate coefficient. {\it
Triangles}, total rate in LTE, summed over all initial states, from the
cross sections by Takagi ~(2002); {\it dashed curve}, fit by GP98 based
on data by Schneider et al.~(1994), given in the temperature range
20~K$<T_{\rm gas}<4000$~K; {\it solid curve}, fit
reported in Table \ref{fits}.}
\label{takagifitcs} 
\end{figure}

Different studies have focused on DE and/or DR of H$_+^2$ (e.g., Takagi~2002; Stroe \&
Fifirig~2009; Fifirig \& Stroe 2008; Motapon et al.~2008; Takagi et al.~2009).
 We have used data from the work by Takagi~(2002), where
DR and DE cross sections have been calculated using the
\emph{multi-quantum defect theory} (MQDT). For each vibrational level,
Takagi~(2002) provided both DR and DE cross sections.
All excited dissociative states are included (all
Rydberg states of 5 different symmetries) and for the lowest
dissociative state, the electronic interaction is fully taken into
account. Fig.~\ref{takagifitcs} shows the comparison between the total
dissociative recombination rate coefficient calculated in the
hypothesis of LTE using the cross sections by Takagi~(2002), and the
fit by GP98 based on data by Schneider et al.~(1994).  Our fit is given
in Table~\ref{fits}.  Although the cross sections of this process have
a threshold component which corresponds to a repulsive channel, the
threshold feature is located at too high energy to affect results in
the present application.  The main features relevant here are instead
the presence of many resonances at low energy which convolute
 into a strongly decreasing trend for energies below a few eV, and a
 global increase of this low energy portion of the cross sections with
$v$. As a result, the LTE rate coefficient decreases with
$T_{\rm gas}$ until a few thousand K, where the
raising population of vibrational levels reverses this trend. The fit
 by GP98, reported in Fig.~\ref{takagifitcs} for reference,
was based on data by Schneider et al.~(1994), given 
up to $T_{\rm gas}=4000$~K and
therefore showed a monotonic trend.

\subsection{Dissociative attachment}

Electron-impact inelastic processes of vibrationally excited 
H$_2$ molecules,
\[
\mathrm{H_2}(v)+\mathrm{e}^-\rightarrow \mathrm{H}+\mathrm{H^-},
\]
play an important role in the kinetics of a low-temperature hydrogen
plasma. Especially for inelastic collisions involving an electronic
transition, two important features can be noted: 
if a vibrationally excited molecule is involved, the threshold
decreases and the cross section of the process
increases (for some processes dramatically) with increasing vibrational
excitation of the molecule.  In this work we have used
data calculated by Celiberto et al.~(2001), based on an improved
version of the resonant scattering model originally developed by
Fano~(1961). The dynamics of this route provides for the formation of a
temporary negative molecular ion, whose fate is either ionization or
dissociation. The fit of the LTE rate coefficient can be found in
Capitelli et al.~(2007). 

\subsection{Electron collisional excitation of $\mathrm{H_2}$}

The process of vibrational excitation of molecules by
electron collisions, 
\[
\mathrm{H_2}(v)+\mathrm{e^-} \rightarrow \mathrm{H_2}(v^\prime)+\mathrm{e^-}+h\nu,
\]
also known as the E-V process, selectively populates high
vibrational levels and represents one of the main non-equilibrium pathways
in the vibrational kinetics of hydrogen plasma. It is a two-step process
that links a vibrational state of the electronic ground state to another
vibrational state of the same manifold, via an intermediate singlet
state that can radiatively decay. As in the case of
dissociative attachment, we have used data from Celiberto
et al.~(2001).

\subsection{Formation of $\mathrm{H_2^+}$ via $\mathrm{HeH^+}$}

The $\mathrm{HeH^+}$ channel for the formation of $\mathrm{H_2^+}(v)$,
\[
\mathrm{HeH^+} + \mathrm{H} \rightarrow \mathrm{H_2^+}(v)+\mathrm{He},
\]
has been considered in several studies (e.g. Dalgarno~2005, HP06,
S08).  The rate coefficient usually adopted in these kinetics models
is estimated from the experimental work by Karpas et
al.~(1979) or data by Linder et al.~(1995). However,
no vibrationally resolved information is available. In order
to overcome this limitation, we follow here an approach
similar to that of Hirata \& Padmanabhan~(2006), where this reaction is
assumed to equally populate all levels compatible with energy balance,
i.e. all levels with energy $E<-1.844$~eV with respect to the H$_2^+$
dissociation limits. Since in the present work individual
rovibrational levels are not considered, we follow
a simplified approach by assuming that this reaction channel produces
H$_2^+$ molecules in the first three vibrational levels with equal rates.

\subsection{Radiative transitions of $\mathrm{H_2}$}

Quadrupolar transition probabilities for
$\mathrm{H_2}(v)$ have been calculated by Wolniewicz et
al.~(1998)\footnote{Data in electronic format are avalaible at
{\tt http://cfa-www.harvard.edu/$^\sim$simbotin/4pole.html}.}.
The authors improved earlier calculations by Turner
et al.~(1977) by adopting a more accurate potential
energy curve, providing tables for the spontaneous decay from an
initial to a final vibrational level for $\Delta
J=0,\pm 2$. The rovibrational energy levels used to perform the sum
of available transition rates have been calculated
using the WKB method (Esposito~2010, priv. comm.). They are in very
good agreement with classical calculations by Kolos \&
Wolniewicz~(1964).

\subsection{Radiative transitions of $\mathrm{H_2^+}$}

Calculations of quadrupole transition probabilities for H$_2^+(v)$
follow the method used by Wolniewicz et al.~(1998)
for H$_2$. Radiative transitions for H$_2^+$
have been calculated by Posen et al.~(1983) for
vibrational-vibrational transitions at given $J$ and $\Delta J$.
The energies of the rovibrational levels used to perform
the sum have been calculated by Hunter et al.~(1974).

\section{Chemical network} 
\label{bozomath}

To describe the time evolution of the chemical species during the
expansion of the Universe, we solve numerically
the system of ODEs 
\beq
\begin{split}
\frac{dn_v}{dt}=&-n_v\sum_{v'}(R_{vv'}+P_{vv'}+C_{vv'}n_{v'})+\\
&+\sum_{v'}R_{vv'}n_{v'}+\sum_{v'}\sum_{v''}\mathbf{C}^{v'v''}_vn_{v'}n_{v''}. \\
\end{split}
\label{ode}
\eeq
where $R_{vv^\prime}$ are the spontaneous and stimulated
radiative rate coefficients; $P_{vv^\prime}$ are the
destruction rate coefficients of the $v$-th species by photons; $C_{vv^\prime}$ are the destruction rate
coefficients for the $v$-th species for collisions
with the $v^\prime$-th chemical partner; and ${\bf
C}^{v^\prime v^{\prime\prime}}_v$ are the formation
rate coefficients of the $v$-th species due to collisions between the
$v^\prime$-th and $v^{\prime\prime}$-th species, photons included.

The radiative rate coefficients are 
related to the Einstein coefficients by
\[
\begin{split}
R_{vv^\prime} = & \left\{
\begin{array}{ll}
A_{vv^\prime}B_{vv^\prime}u(\nu_{vv^\prime},T_{\rm rad}) & \mbox{if $v^\prime<v$} \\
B_{vv^\prime}u(\nu_{vv^\prime},T_{\rm rad}) & \mbox{if $v^\prime>v$,} \\
\end{array} \right. \\
= & \left\{
\begin{array}{ll}
A_{vv^\prime}[1+\eta(\nu_{vv^\prime},T_{\rm rad})] & \mbox{if $v^\prime<v$} \\
g_{v^\prime}A_{v^\prime v}\eta(\nu_{vv^\prime},T_{\rm rad})/g_v & \mbox{if $v^\prime>v$} \\
\end{array} \right.
\end{split}
\]
where $\nu_{vv^\prime}$ is the frequency of the transition $v\rightarrow
v^\prime$, $g_v=1$ is statistical weight of the $v$-th level,
$u(\nu_{vv^\prime},T_{\rm rad})$ is the Planck photon distribution,
and $\eta(\nu_{vv^\prime},T_{\rm rad})=[\exp(h\nu_{vv^\prime}/kT_{\rm
rad})-1]^{-1}$.  

The chemical network is completed by the 
equations for the gas and radiation temperature evolutions
and the equation for the redshift,
\beq
\frac{dt}{dz}=-\frac{1}{(1+z)H(z)}, 
\eeq
where
\beq
H(z)=H_0 \sqrt{\Omega_{\rm r}(1+z)^4+\Omega_{\rm m}(1+z)^3+\Omega_{\rm K}(1+z)^2+\Omega_\Lambda}.
\eeq
The number density of hydrogen atoms is 
\beq
n_{\rm H}=\Omega_b \frac{3 H_0^2}{8\pi G m_{\rm H}}(1-Y_p)(1+z)^3,
\label{nb}
\eeq
where $\Omega_b$ is the baryon fraction, $Y_p$ is the helium mass
fraction, $G$ is the constant of gravitation and $m_{\rm H}$ the atomic hydrogen mass.


Numerical values of the cosmological parameters $H_0$, $T_0$, $\Omega_r$,
$\Omega_m$, $\Omega_b$, $\Omega_\Lambda$ and $Y_p$ have been 
obtained
from WMAP5 data (Komatsu et al.~2009, see Table~\ref{tablecosmo}). The initial fractional abundances for 
H, He and D are also listed in the same table. The hydrogen
ionization fraction was computed using the routine RECFAST (Seager et
al.~1999, Seager et al.~2000, Wong et al.~2008). Helium
was assumed to be fully neutral in the redshift range of interest
($z\le 2000$). The electron density was determined by imposing
the condition of charge neutrality.  Atomic and molecular weights
were taken from the NIST webpage \footnote{\tt
http://webbook.nist.gov/chemistry/name-ser.html}.

\begin{table}[ht]
\caption{Cosmological parameters}
\begin{tabular*}{\columnwidth}{ll}
\hline
Cosmological parameter & Numerical value\\
\hline 
$H_0$ & $100\,h$~km~s$^{-1}$~Mpc$^{-1}$ \\
$h$ & $0.705$\\
$z_{\rm eq}$ & $3141$\\
$T_0$ & $2.725$~K\\
$\Omega_{\rm dm}$ & $0.228$\\
$\Omega_{\rm b}$ & $0.0456$\\
$\Omega_{\rm m}$ & $\Omega_{\rm dm}+\Omega_{\rm b}$\\
$\Omega_{\rm r}$ & $\Omega_{m}/(1+z_{\rm eq})$\\
$\Omega_\Lambda$ & $0.726$\\
$\Omega_{\rm K}$ & $1-\Omega_{\rm r}-\Omega_{\rm m}-\Omega_\Lambda$ \\
$Y_p$ & $0.24$\\
$f_{\rm H}$ &  0.924\\
$f_{\rm He}$ & 0.076\\
$f_{\rm D}$ &  $2.38\times 10^{-5}$\\      
& \\
\hline
\end{tabular*}
\label{tablecosmo}          
\end{table}



Data tables of vibrationally resolved rate coefficients
have been inserted as input of the routine LSODE \footnote{\tt
https://computation.llnl.gov/casc/odepack/download/lsode\_agree.html}
used to integrate the chemical system, which consists of 47 differential
equations, one for each chemical species introduced in the model and one
for the gas temperature.  An implicit method has been used to perform
the integration, being the kinetic problem stiff. Linear interpolation
of the rate coefficients and of the ionization fraction is performed in
logarithmic scale at each step of integration.

\section{Results and discussion}
\begin{figure}
\includegraphics[width=8cm]{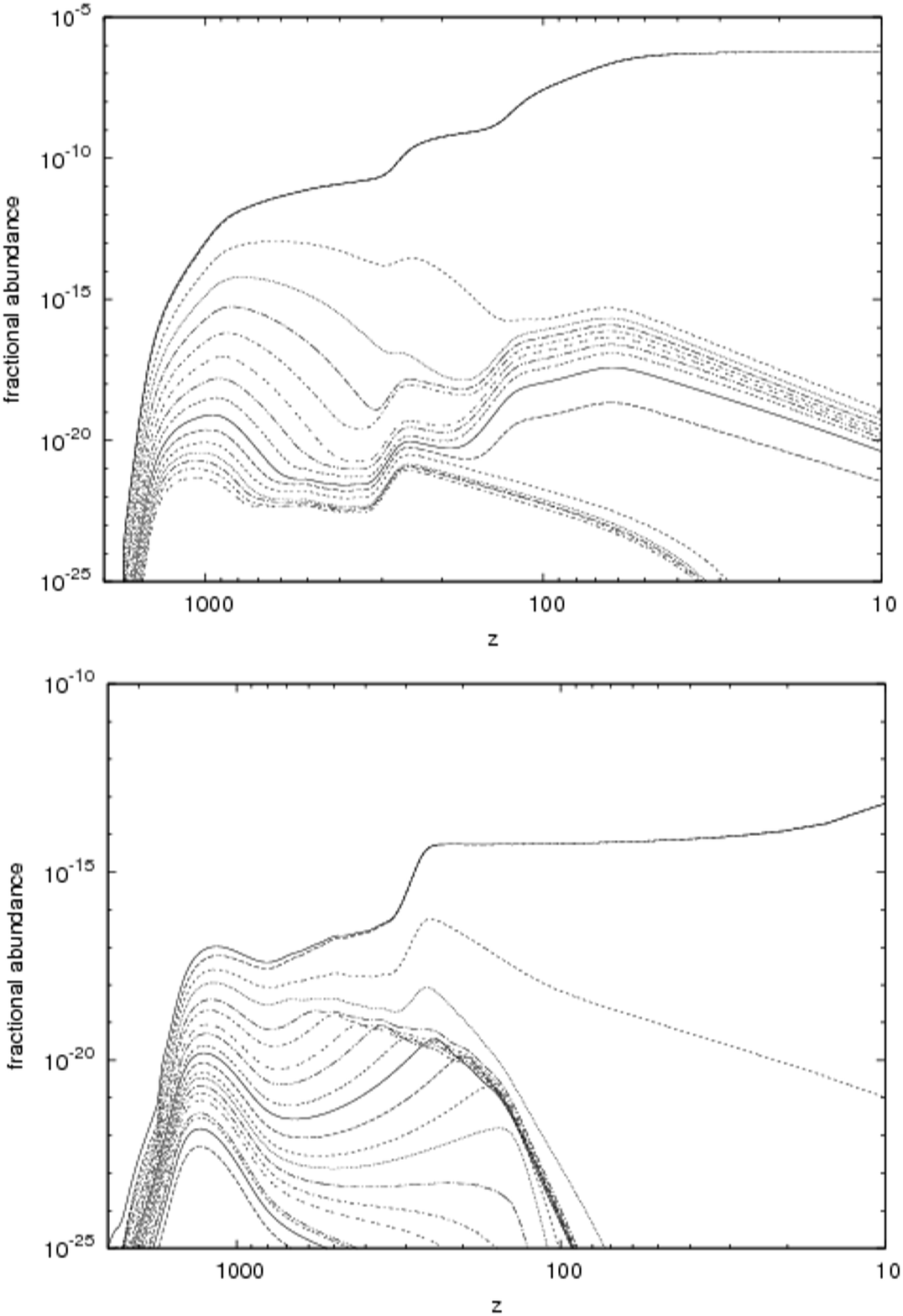} 
\caption{Vibrational level populations of H$_2$ ({\it top panel}, from
$v=0$ to $v=14$) and H$_2^+$ ({\it bottom panel}, from $v=0$ to $v=18$)
as function of redshift.  In both cases, the {\it solid curve} shows
the total fractional abundance.}
\label{case0} 
\end{figure}

The resulting fractional abundances of $\mathrm{H_2}$ and
$\mathrm{H_2^+}$ for each vibrational level are shown as
a function of redshift in Fig.~\ref{case0}. The total
fractional abundance of H$_2$ shows a rapid increase at three epochs:
at $z\approx 1500$ by the charge transfer channel,
dominant at high temperatures, at $z\approx 300$
by H$_2^+$ radiative association formation, that modulates the
charge transfer channel, and at $z\approx 100$ by the
associative detachment process.
At $z=10$ the fractional abundances of
H$_2$ and H$_2^+$ are $5.76\times 10^{-7}$ and $6.56\times 10^{-14}$, respectively.
Our value of the H$_2$ abundance is in good agreement with that obtained by
HP06 ($f({\rm H}_2)=6.0\times 10^{-7}$ at $z=20$) but about twice the value
of V09 ($2.7\times 10^{-7}$ at $z=10$).
As for H$_2^+$, our abundance is in good agreement with the result of V09
($f({\rm H}_2^+)=6.7\times 10^{-14}$ at $z=10$), whereas HP06 obtain three
different abundances depending on the value of the H$_2^+$--H charge exchange reaction.

For H$_2$, a marked non-equilibrium distribution of
populations is established at $z\approx 300$, followed
by a plateau involving levels from $v=1$ to $9$. Levels above $v=9$
are excluded from this last plateau as their populations
drop dramatically due to the endoergic character of
the process of associative detachment (see Section~\ref{subcizek}).
Therefore, when the thermal kinetic energy is much lower than the energy
thresholds, the formation of highly excited ($v\geq10$) $\mathrm{H_2}$
molecules is strongly suppressed.  Such a strong splitting of level
histories is not observed at higher $z$, since the main H$_2$ formation
mechanism  ($\mathrm{H+H_2^+}$ charge transfer) is exoergic for all $v$
(in constrast with the charge transfer $\mathrm{H^++H_2}$, that is a
threshold process).

\begin{figure}
\includegraphics[width=8cm]{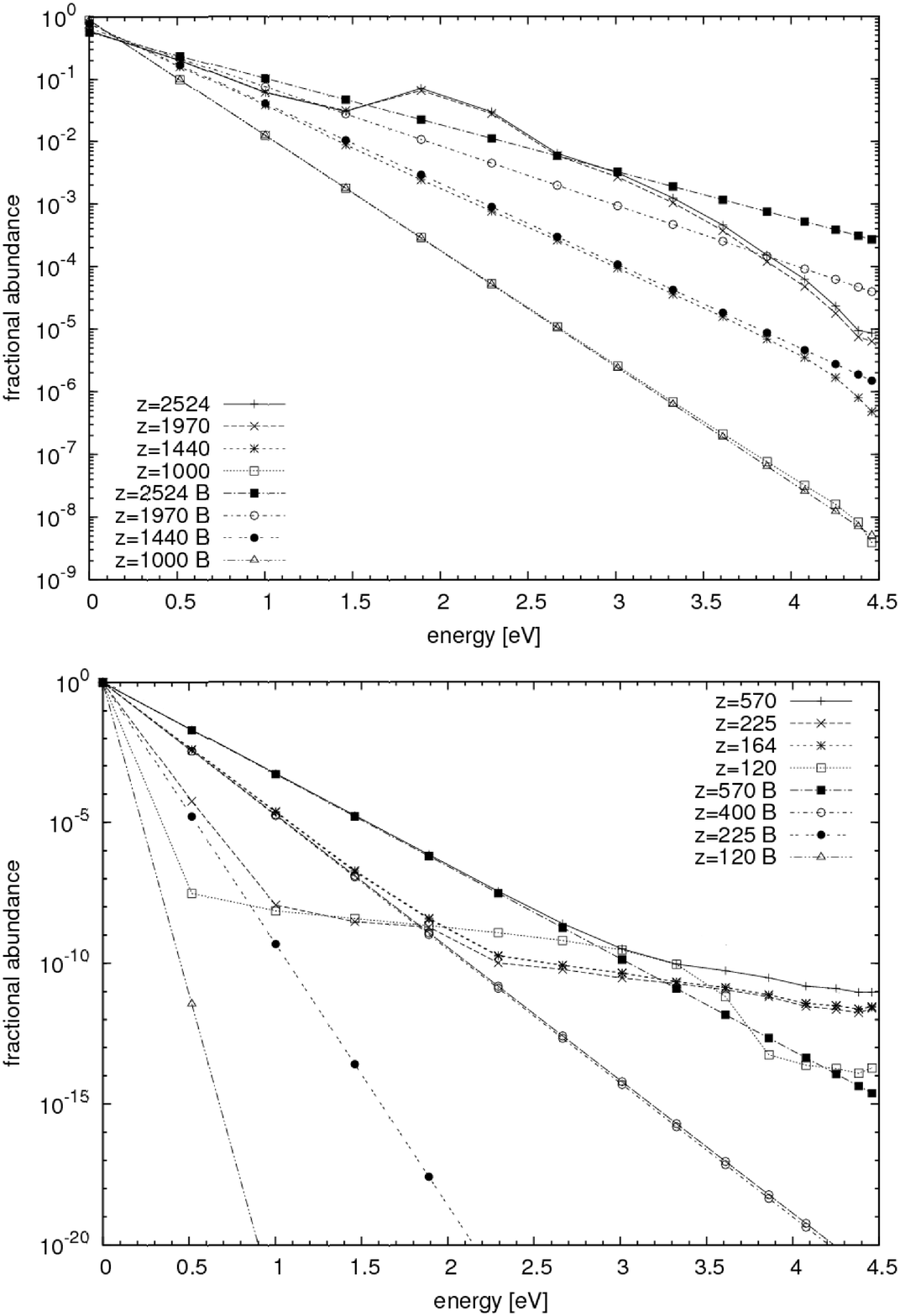} 
\caption{Vibrational level populations of H$_2$, normalized to the
total H$_2$ fractional abundance, at various redshifts.  The Boltzmann
distributions at each redshift are indicated with the label ``B''.}
\label{vdfcase0} 
\end{figure}

The $\mathrm{H_2}$ vibrational distribution functions obtained at
different values of the redshift are shown in Fig.~\ref{vdfcase0},
compared with the equilibrium curves corresponding to the
Boltzmann vibrational distributions at the corresponding
value of the gas temperatures. The shape of the level population
distribution of H$_2$ at low $z$ can be understood and even roughly
evaluated on the basis of a balance between radiative and formation rates.
Indeed, the bunching of population of excited levels for low $z$ is due to 
the fact that each population is determined essentially from the ratio 
of associative detachment and radiative rates.
The ratios for all exothermic channels ($v<$10) have comparable values
 on a enlarged logarithmic scale like that used here.

\begin{figure}
\includegraphics[width=8cm]{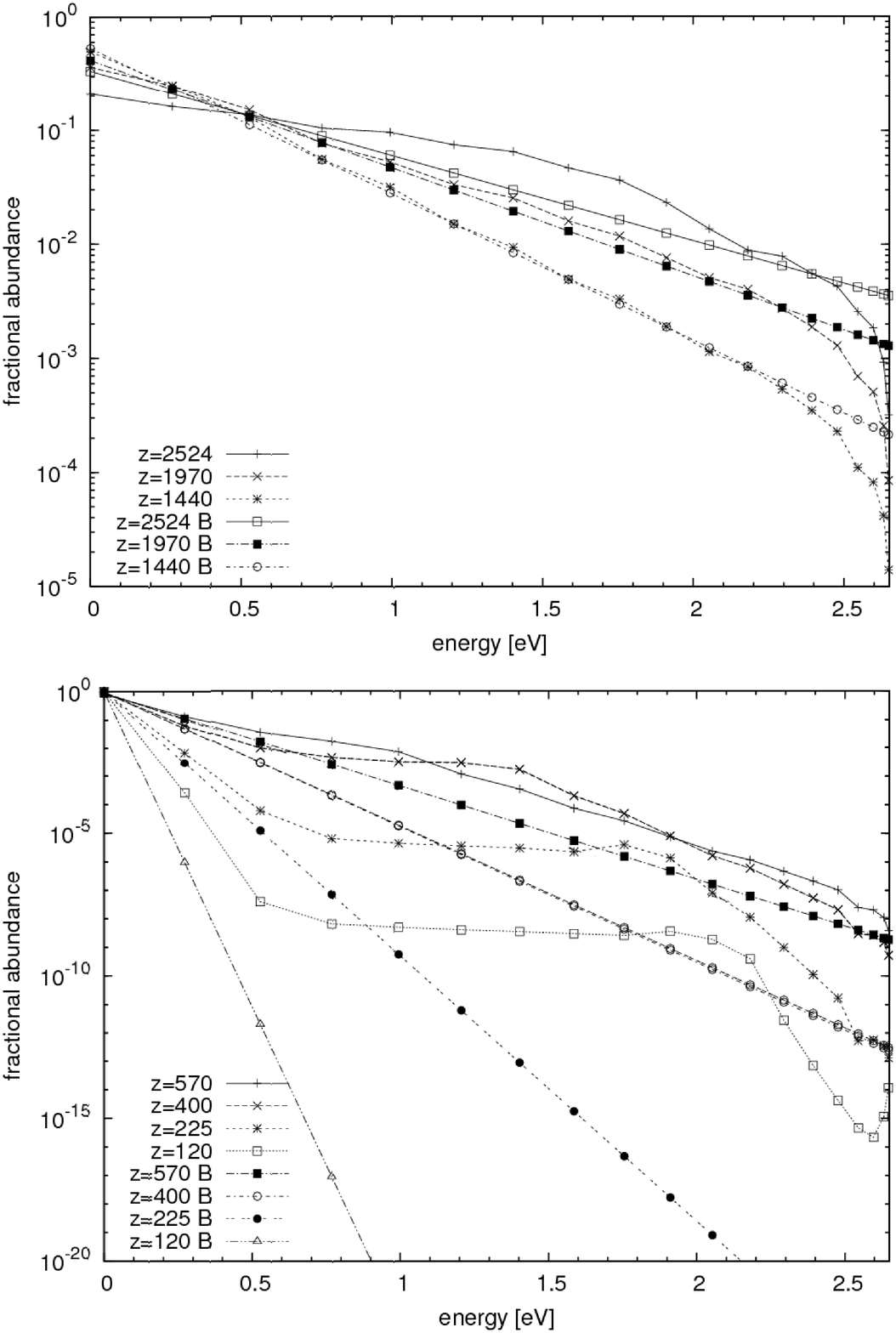} 
\caption{Same as Fig.~\ref{vdfcase0} for H$_2^+$.}
\label{vdfh2pcase0} 
\end{figure}

For H$_2^+$, the most relevant formation process is radiative
association. The vibrational resolution of this species shows that
the formation process, being faster for highly excited vibrational
levels of the products, leads to a pronounced tail in the vibrational
distribution (Fig. \ref{vdfh2pcase0}). The latter is characterized by a
high population of the first few vibrational levels followed by a
suprathermal tail, in qualitative agreement with the
results of HP06 (see their Fig.~4). However, for HP06 the dominant
formation process of H$_2^+$ is via HeH$^+$. This
result supports the claim of HP06 that the radiative
association channel plays a minor role among formation pathways in their
model. Also in this case, we observe a strong grouping of the fractional
abundance of excited levels which is explained by the balance between
state specific level production rates and radiative
transitions to $v=0$.

\begin{figure}
\includegraphics[width=8cm]{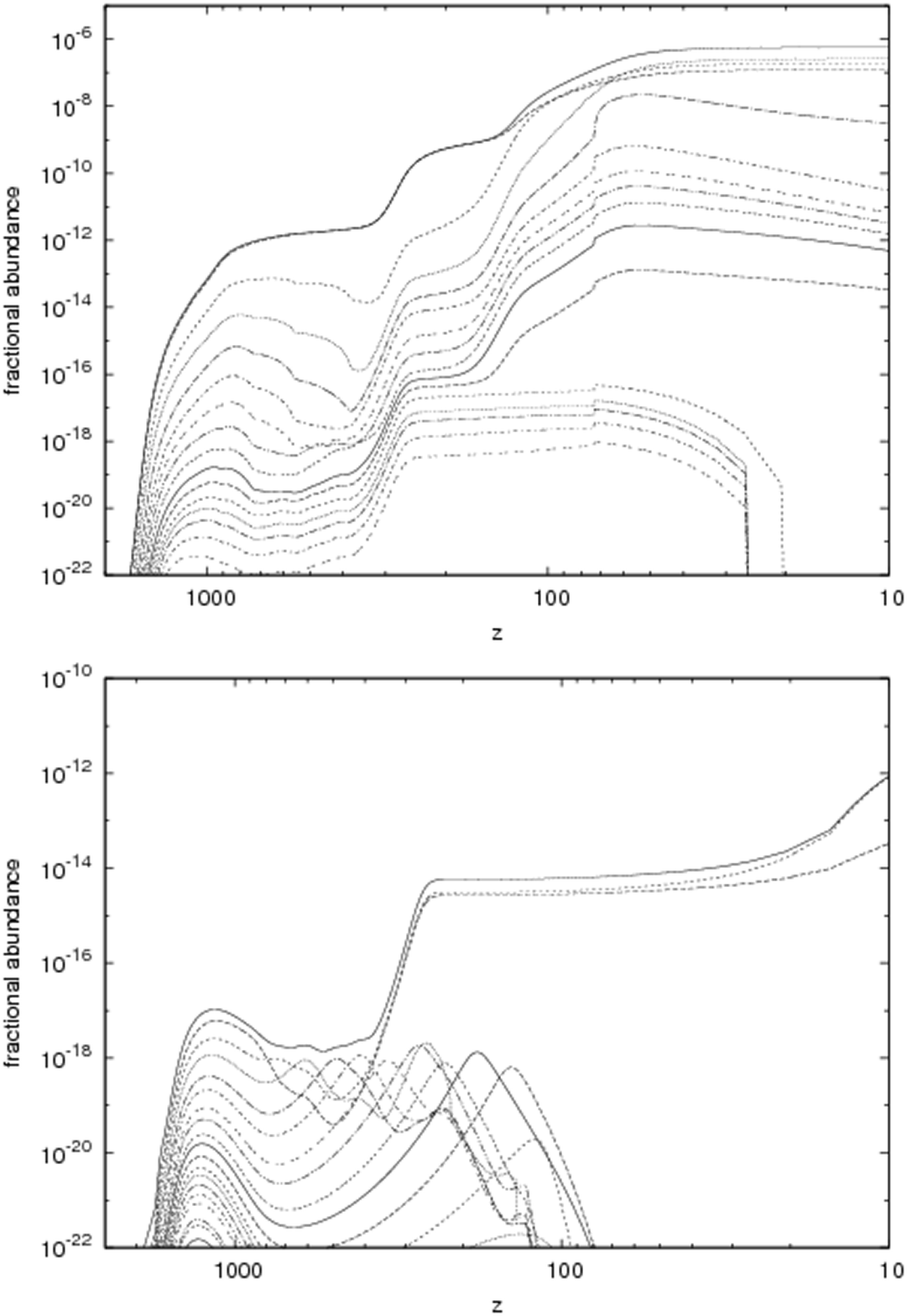}
\caption{Same as Fig.~\ref{case0}, including formation
and destruction channels, V-T processes but no radiative transition.}
\label{case1}
\end{figure}

These findings suggest that the processes leading to redistribution
of the vibrational quanta need further attention. This study is also appropriate
here because such processes cannot be modeled in the
framework of the usual, not state-resolved approach. In our case,
such redistributing channels are V-T and spontaneous/stimulated
radiative processes.  In order to understand how these pathways effect
the vibrational distribution of H$_2$ and H$_2^+$, further numerical
experiments have been performed in various regimes,
corresponding to particular physical conditions. Fig.~\ref{case1}
shows the results obtained for the case in which radiative processes
have been omitted.  The figure shows that, although
radiative processes are not fast enough to thermalize level populations,
they are essential to produce the actual vibrational distribution,
and their neglect leads to complete different results
especially at low $z$.

\begin{figure}
\includegraphics[width=8cm]{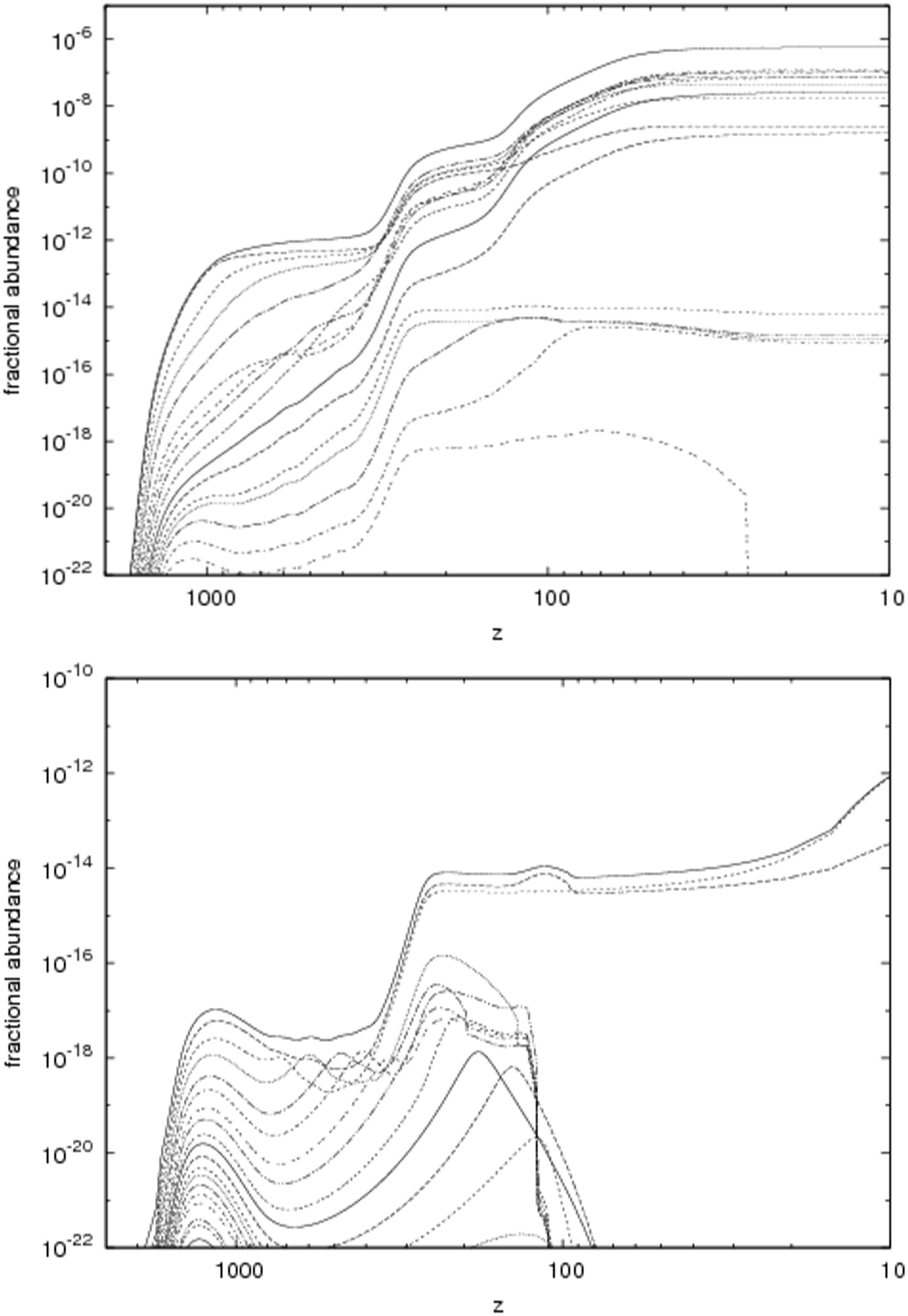}
\caption{Same as Fig.~\ref{case0}, including formation
and destruction channels, but no V-T process and radiative transition.}
\label{case3}
\end{figure}

On the other hand, removal of V-T processes does
not produce any appreciable variation with respect to the results of
Fig.~\ref{case0}, since the rates of V-T processes
are much smaller than the radiative ones. The role of V-T processes
is better appreciated by looking at Fig.~\ref{case3}, where
V-T processes and radiative decay are both ignored. Some variations with respect to Fig.~\ref{case1} are observed: a higher
vibrational temperature, a different high $v$ plateau for H$_2$,
especially at low $z$, and some differences of the total H$_2^+$ 
fraction. 

The main destruction channel (dissociative
recombination) active at lower $z$ is deeply affected by radiative
processes: when these are ignored,
the destruction channel becomes more efficient and the
fractional abundance at the second peak (at $z \sim 300$) is reduced by
about a factor of 2. As a further
check, removing radiative processes together with the loss channel leads
to small variations of the fractional abundance.
Thus, redistribution among levels is able
to affect even the results for the total $\mathrm{H_2^+}$, confirming
the relevance of the non-equilibrium vibrational distributions.

No rotationally resolved data exists for most of the formation and destruction processes included
in the present work, so a state-to-state kinetics cannot at the present stage go beyond the resolution
of molecular vibrations. For H$_2$ and H$_2^+$, the typical rotational energies are of the order of 0.01 eV:
this suggests that the way in which rotation is included in the kinetic model can affect the
low $z$ trend of molecular abundance.
To test the sensitivity of the chemical abundances to the rotational structure of H$_2$ and H$_2^+$,
we have computed the H$_2^+$ radiative association rate coefficient (Subsection \ref{radasssection}) assuming that
the rotational manifold for each vibrational level reduces to only one J, either the lowest or the highest. 
 Even if this hypothesis implies a very strong rotational 
non-equilibrium, it has negligible effect on the freeze-out
value of H$_2^+$, although the abundance of this species at $z \simeq 1000$ is changed. Furthermore,
 since the photodissociation
 cross section is not rotationally resolved, the rotational degree of freedom only enters in Eq.(\ref{rcradass}) through $Z_{rot}$.
 As a consequence, the nascent vibrational distribution function, which is the most important quantity here, is only marginally modified.

\section{Conclusions}

In this work we have performed calculations of the vibrational
distribution of both H$_2$ and H$_2^+$ for the conditions expected in
the early Universe and based on a comprehensive state-to-state chemical
kinetics. The vibrational level distribution for these two species are
reported for a wide and continuous range of the redshift parameter.
The results can be summarized as follows:

\begin{enumerate}

\item The vibrational distribution function of H$_2$ and H$_2^+$ assumes a
quasi-equilibrium shape at redshift $z \sim 1500$; after that, extended
plateau in the vibrational level distributions form, underlying the
presence of pumping phenomena for the intermediate vibrational levels;
full thermalization is not observed, because vibrational relaxation
processes are not fast enough to balance the strong vibrational
selectivity of formation rates.

\item Radiative processes play a fundamental role in the redistribution of
vibrational quanta, affecting the vibrational distribution function and
the overall fractional abundance (as in the case of H$_2^+$);

\item All these features can not be described in terms of the LTE
distributions which are usually assumed in chemical networks for the
primordial Universe.

\end{enumerate} 

\acknowledgments

We are grateful to the following colleagues for their precious aid in
retrieving vibrationally resolved cross sections: Roberto Celiberto,
Martin \u{C}\'{\i}\v{z}ek, Gordon Dunn, Fabrizio Esposito, Predrag
Krsti\'{c}, Ratko Janev, Ralph Jaquet, Susanta Mahapatra, Donald
Shemansky, Hidekazu Takagi, Xavier Urbain. C.M.C. would also acknowledge Jonathan Tennyson and Ioan F. Schneider for useful discussions on electron$-$molecule collisions, and MIUR
 and Universit\`{a} degli Studi di Bari, that partially supported this work (\textquotedblleft fondi di Ateneo 2010\textquotedblright).

\appendix

\section{Analytical fits of LTE rate coefficients}

In the present section, the analytic expressions of the vibrational LTE
rate coefficients of the chemical processes introduced in the present
model are reported. For processes that are vibrationally
resolved in the final states, the sum on the entire final manifold has
also been carried out.

\begin{table}[ht]
\caption{Chemical reactions and their LTE rate coefficients}
\begin{tabular*}{\columnwidth}{ll}
\hline
Chemical process & Rate coefficient (m$^3$~s$^{-1}$) or (s$^{-1}$) \\
\hline
& \\
$1] ~~\mathrm{H}+\mathrm{H^-} \rightarrow \mathrm{H_2}(v) + \mathrm{e^-}$  & 
$\log_{10}k=\sum_{n=0}^2 a_{2n}[\log_{10}(T_{\rm gas}/10^2)]^{2n}$ \\
  & $a_0=-14.4$ \\
  & $a_2=-0.15$ \\
  & $a_4=-7.9\times 10^{-3}$\\
& \\
$2] ~~\mathrm{H_2}(v)+\mathrm{H^+} \rightarrow \mathrm{H_2^+}(v^\prime)+\mathrm{H}$ & 
$\ln k=a_0+a_1 T_{\rm gas}+a_2/T_{\rm gas}+a_3 T_{\rm gas}^2$, \\
  & $a_0=-33.081$\\
  & $a_1=6.3173\times 10^{-5}$\\
  & $a_2=-2.3478\times 10^4$\\
  & $a_3=-1.8691\times 10^{-9}$\\
& \\
$3] ~~\mathrm{H_2^+}(v)+\mathrm{H} \rightarrow \mathrm{H^+}+\mathrm{H}+\mathrm{H}$  & 
$\ln k=a_0+a_1 T_{\rm gas}+a_2/T_{\rm gas}+a_3 T_{\rm gas}^2$, \\
  & $a_0=-32.912$\\
  & $a_1=6.9498\times 10^{-5}$\\
  & $a_2=-3.3248\times 10^4$\\
  & $a_3=-4.08\times 10^{-9}$\\
& \\
$4] ~~\mathrm{H_2}(v)+\mathrm{H^+} \rightarrow \mathrm{H}+\mathrm{H}+\mathrm{H^+}$ &  
$\ln k=a_0+a_1 T_{\rm gas}+a_2/T_{\rm gas}+a_3 T_{\rm gas}^2$, \\
  & $a_0=-33.404$\\
  & $a_1=2.0148\times 10^{-4}$\\
  & $a_2=-5.2674 \times 10^4$\\
  & $a_3=-1.0196\times 10^{-8}$\\
& \\
$5] ~~\mathrm{H_2}(v)+\mathrm{H} \rightarrow \mathrm{H}+\mathrm{H}+\mathrm{H}$ & 
$k=1.9535\times 10^{-10} T_{\rm gas}^{-0.93267}\exp(-4.9743\times 10^4/T_{\rm gas})$ \\
& \\
$6] ~~\mathrm{H_2}(v)+h\nu \rightarrow \mathrm{H_2^+}(v^\prime)+\mathrm{e^-}$ & 
$k=3.06587\times 10^9 \exp(-1.89481\times 10^5/T_{\rm gas})$ \\
& \\
& \\
$7] ~~\mathrm{H_2}(v)+h\nu \rightarrow \mathrm{H}+\mathrm{H}$ (indirect)& 
$\ln k={17.555+7.2643\times 10^{-6}~T_{\rm gas}-1.4194\times 10^5/T_{\rm gas}}$ \\
& \\
$8] ~~\mathrm{H_2^+}(v)+e^- \rightarrow 2\mathrm{H}$ & 
$k=\sum_{n=0}^5 a_n T_{\rm gas}^n$, \\
  & $a_0=4.2278\times 10^{-14}$\\
  & $a_1=-2.3088\times 10^{-17}$\\
  & $a_2=7.3428\times 10^{-21}$\\
  & $a_3=-7.5474\times 10^{-25}$\\
  & $a_4=3.3468\times 10^{-29}$\\
  & $a_5=-5.528\times 10^{-34}$\\
& \\
$9] ~~\mathrm{H^-}+h\nu \rightarrow \mathrm{H}+\mathrm{e^-}$ & 
    $k=0.11 T_{\rm rad}^{2.13}\exp(-8.823\times 10^3/T_{\rm rad})$ \\
  & $k=8.0\times 10^{-8}T_{\rm rad}^{1.3}\exp(-2.3\times 10^3/T_{\rm rad})$ (non-thermal)\\
& \\
\hline
\end{tabular*}
\label{fits}          
\end{table}

\end{document}